\documentclass[9pt,twocolumn,twoside]{gsajnl}

\usepackage[utf8]{inputenc}
\usepackage[T1]{fontenc}
\usepackage{amsfonts}
\usepackage{nicefrac}
\usepackage{amsmath}
\usepackage[numbers]{natbib}
\usepackage{hyperref}
\hypersetup{
    colorlinks=true,
    linkcolor=blue,
    filecolor=magenta,      
    urlcolor=cyan,
    pdftitle={Choroidalyzer},
    pdfpagemode=FullScreen,
    }
\usepackage[capitalise]{cleveref}
\usepackage{placeins}

\usepackage{graphicx}
\usepackage{booktabs}

\usepackage{caption}        
\usepackage{subcaption}     
\usepackage{multirow}
\usepackage{xcolor}

\usepackage{microtype}
\usepackage{contour}
\usepackage{adjustbox}

\newcommand{\absdiv}[1]{%
  \par\addvspace{.5\baselineskip}
  \noindent\textbf{#1}\quad\ignorespaces
}

\renewenvironment{thebibliography}[1]{
  \begin{oldthebibliography}{#1}
    \setlength{\itemsep}{0em}
    \setlength{\parskip}{0em}
}
{
  \end{oldthebibliography}
}

\title{Choroidalyzer: An open-source, end-to-end pipeline for choroidal analysis in optical coherence tomography}
\runningtitle{Choroidalyzer: End-to-end choroidal analysis}
\runningauthor{Engelmann, Burke, et al.}

\author[$\ast$,1,2]{Justin Engelmann}
\author[$\ast$,3]{Jamie Burke}
\author[4]{Charlene Hamid}
\author[4]{Megan Reid-Schachter}
\author[5]{Dan Pugh}
\author[5]{Neeraj Dhaun}
\author[6]{Diana Moukaddem}
\author[6]{Lyle Gray}
\author[6]{Niall Strang}
\author[7]{Paul McGraw}
\author[8]{Amos Storkey}
\author[9]{Paul J. Steptoe}
\author[3]{Stuart King}
\author[4,10]{Tom MacGillivray}
\author[2,11]{Miguel O. Bernabeu}
\author[8,10]{Ian J.C. MacCormick}

\affil[1]{School of Informatics, University of Edinburgh, Edinburgh, UK}
\affil[2]{Centre for Medical Informatics, University of Edinburgh, Edinburgh, UK}
\affil[3]{School of Mathematics, University of Edinburgh, Edinburgh, UK}
\affil[4]{Clinical Research Facility and Imaging, University of Edinburgh, Edinburgh, UK}
\affil[5]{British Heart Foundation Centre for Cardiovascular Science, University of Edinburgh, Edinburgh, UK}
\affil[6]{Department of Vision Sciences, Glasgow Caledonian University, Glasgow, UK}
\affil[7]{School of Psychology, University of Nottingham, Nottingham, UK}
\affil[8]{Institute for Adaptive and Neural Computation, School of Informatics, University of Edinburgh, Edinburgh, UK}
\affil[9]{Princess Alexandra Eye Pavilion, NHS Lothian, Edinburgh, UK}
\affil[10]{Centre for Clinical Brain Sciences, University of Edinburgh, Edinburgh, UK }
\affil[11]{The Bayes Centre, University of Edinburgh, Edinburgh, UK}

\correspondingauthoraffiliation[$\ast$]{Corresponding and lead authors with equal contribution; \\ Email address:  Justin.Engelmann@ed.ac.uk \& Jamie.Burke@ed.ac.uk; \\ Word count: na; \\ Funding information: UK Research and Innovation [grant number EP/S02431X/1] and Medical Research Council [grant number MR/N013166/1]; \\ Commercial relationships: \textbf{Justin Engelmann}, None; \textbf{Jamie Burke}, None; \textbf{Charlene Hamid}, None; \textbf{Megan Reid-Schachter}, None; \textbf{Tom Pearson}, None; \textbf{Dan Pugh}, None; \textbf{Neeraj Dhaun}, None; \textbf{Diana Moukaddem}, None; \textbf{Lyle Gray}, None; \textbf{Paul McGraw}, None; \textbf{Niall Strang}, None; \textbf{Amos Storkey}, None; \textbf{Paul J. Steptoe}, None; \textbf{Stuart King}, None; \textbf{Tom MacGillivray}, None; \textbf{Miguel O. Bernabeu}, None;  \textbf{Ian J.C. MacCormick}, None.}

\begin{abstract}
\absdiv{Purpose} 
To develop Choroidalyzer, an open-source, end-to-end pipeline for segmenting the choroid region, vessels, and fovea, and deriving choroidal thickness, area, and vascular index.

\absdiv{Methods}
We used 5,600 OCT B-scans (233 subjects, 6 systemic disease cohorts, 3 device types, 2 manufacturers). To generate region and vessel ground-truths, we used state-of-the-art automatic methods following manual correction of inaccurate segmentations, with foveal positions manually annotated. We trained a U-Net deep-learning model to detect the region, vessels, and fovea to calculate choroid thickness, area, and vascular index in a fovea-centred region of interest. We analysed segmentation agreement (AUC, Dice) and choroid metrics agreement (Pearson, Spearman, mean absolute error (MAE)) in internal and external test sets. We compared Choroidalyzer to two manual graders on a small subset of external test images and examined cases of high error.

\absdiv{Results} 
Choroidalyzer took 0.299 seconds per image on a standard laptop and achieved excellent region (Dice: internal 0.9789, external 0.9749), very good vessel segmentation performance (Dice: internal 0.8817, external 0.8703) and excellent fovea location prediction (MAE: internal 3.9 pixels, external 3.4 pixels). For thickness, area, and vascular index, Pearson correlations were 0.9754, 0.9815, and 0.8285 (internal) / 0.9831, 0.9779, 0.7948 (external), respectively (all p<0.0001). Choroidalyzer's agreement with graders was comparable to the inter-grader agreement across all metrics. 

\absdiv{Conclusions}
Choroidalyzer is an open-source, end-to-end pipeline that accurately segments the choroid and reliably extracts thickness, area, and vascular index. Especially choroidal vessel segmentation is a difficult and subjective task, and fully-automatic methods like Choroidalyzer could provide objectivity and standardisation. Choroidalyzer is openly available here: \url{https://github.com/justinengelmann/Choroidalyzer}

\end{abstract}

\begin{document}

\maketitle     

\section{Introduction}
The retinal choroid is a densely vascularised tissue at the back of the eye, providing essential nutrients and support to the outer retinal pigment epithelium and photoreceptors \cite{nickla2010multifunctional}. The choroid is emerging as a window into systemic vascular health including brain \cite{robbins2021choroidal}, kidney \cite{balmforth2016chorioretinal}, and heart \cite{yeung2020choroidal}. The choroid is also affected by ophthalmic conditions like myopia \cite{read2019choroidal}. Thus, the choroid is a potential source of biomarkers for ocular and non-ocular disease \cite{burke2023evaluation, burke2023retinal, shin2019evaluation, kundu2023longitudinal}. This is driven by improvements in optical coherence tomography (OCT) imaging, especially enhanced depth imaging OCT (EDI-OCT) \cite{spaide2008enhanced}. Previously, only the retinal layers were well-captured whereas the choroid, which sits below the hyper-reflective retinal pigment epithelium, was not imaged well and thus received little attention. Now, the choroid can be captured well and is a promising frontier for systemic health assessment \cite{tan2016state}, especially as OCT devices become commonplace even at high street optometrists. To compute choroidal metrics that could serve as potential vascular biomarkers like choroidal thickness, area, or vascular index, the choroid region and vasculature must be identified and segmented accurately and reliably. 

While choroidal region segmentation is relatively straightforward compared to vessel segmentation, as only a single shape needs to be identified per scan, accurate detection of the lower choroid boundary (Choroid-Sclera, C-S, junction) can be time consuming and at times ambiguous due to poor contrast or image noise. While semi-automatic methods have been proposed \cite{burke2023evaluation, burke2021edge, eghtedar2022update, masood2018automatic, salafian2018automatic, kajic2012automated, wang2017automatic, srinath2014automated, george2019two, danesh2014segmentation}, these typically require training and expertise to use and do not remove human error and subjectivity. Fully-automatic, deep learning-based approaches to region segmentation have been proposed and address both the time-intensive and the ambiguous nature of region segmentation, drastically improving both the ease and standardisation of choroidal segmentation. Many of these methods are not openly available to the research community \cite{mazzaferri2017open, kugelman2019automatic, devalla2018drunet, chen2022application}, but recently DeepGPET, an open-source choroidal region segmentation method, was published that can be freely downloaded from GitHub \cite{burke2023opensource}.

Choroidal vessel segmentation is a far more complex and time-consuming task. The choroidal vessels are highly heterogeneous in terms of vessel size, shape, and edge contrast and are sometimes hard to discern due to poor contrast or noise, making manual segmentations prohibitively time consuming and very subjective. Currently, local thresholding algorithms are commonplace for choroidal vessel segmentation \cite{branchini2013analysis, sonoda2014choroidal, agrawal2016choroidal}, and the current state-of-the-art is the Niblack algorithm \cite{agrawal2020exploring, betzler2022choroidal}. Niblack is a local thresholding technique which segments the vessels using a fixed-size sliding window and a standard deviation offset to determine a pixel-level threshold. However, there is evidence of wide inter-grader disagreement between the two commonly used adaptations to Niblack's algorithm \cite{wei2018comparison}. Deep learning approaches have been proposed previously trained on manual  annotations or Niblack's algorithm\cite{liu2019robust, muller2022application}, but are not openly available at the time of writing.

Finally, in addition to region and vessel segmentation, there are two more necessary steps that are often overlooked, namely fovea detection and computation of choroidal metrics. OCT B-scans are not necessarily perfectly centred and the size of a pixel can differ not only between devices but also between scans. Thus, once region, vessels, and fovea are extracted, choroidal metrics should be computed in a fovea-centred region-of-interest \cite{burke2023evaluation}, which must account for key details like the pixel-scaling of the scan. Currently, each of these four steps is done by a different tool \cite{zheng2021deep, khaing2021choroidnet} with ad-hoc and non-standardised approaches used especially for fovea detection. \cite{xuan2023deep}.

\begin{figure*}[tb]
\centering
\includegraphics[width=0.95\textwidth]{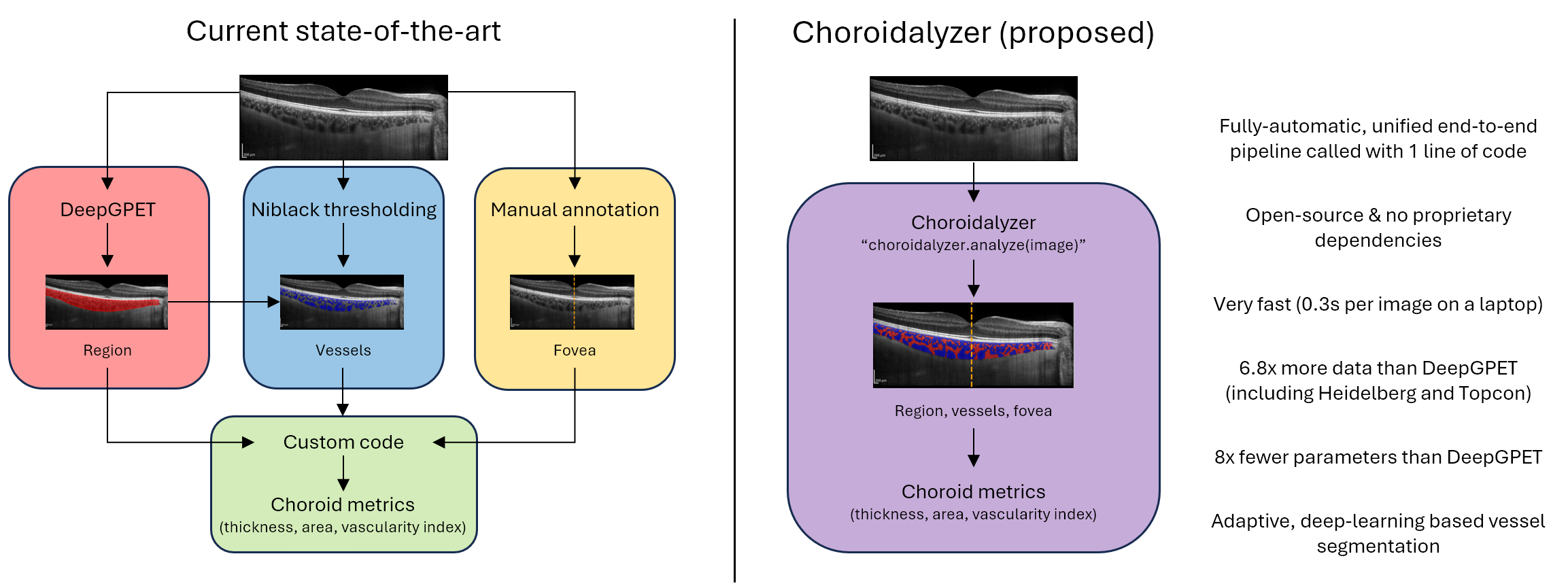}
\caption{A comparison between Choroidalyzer and the existing state of choroidal analysis. To obtain choroidal metrics in a fovea-centred region of interest, researchers currently need to combine many different tools. Choroidalyzer unifies everything into a end-to-end pipeline that is very fast and convenient to use.}
\label{fig:choroidalyzerfig1}
\end{figure*}

We address these issues by proposing Choroidalyzer, an end-to-end pipeline for choroidal analysis. Choroidalyzer consists of a single deep learning model that simultaneously segments the choroidal region and vessels and detects the fovea location, combined with all the code needed to extract choroidal thickness, area, and vascular index in a fovea-centred region of interest. \cref{fig:choroidalyzerfig1} shows how Choroidalyzer improves on the current state-of-the-art by providing a comprehensive solution for all elements of choroidal analysis. To our knowledge, Choroidalyzer is the first open-source method for comprehensive, automatic analysis of the choroid from a raw OCT B-scan. Choroidalyzer is highly effective, can be run on a standard laptop in less than one-third of a second per image, does not require any specialist training in image processing and is available on GitHub: \url{https://github.com/justinengelmann/Choroidalyzer}.

\section{Methods}

\begin{table*}[tb]
\centering
\begin{adjustbox}{max width=\textwidth}
{\small \begin{tabular}{@{}lcccccc|c@{}}
\toprule
 & OCTANE & Diurnal Variation & Normative & i-Test & Prevent Dementia & GCU Topcon & Total \\
  \midrule
\multicolumn{1}{l}{Subjects} & 46 & 20 & 1 & 21 & 121 & 24 & 233\\
\multicolumn{1}{r}{Control/Case} & 0 / 46 & 20 / 0 & 1 / 0 & 11 / 10 & 56 / 65 & 24 / 0 & 112 / 121 \\
\multicolumn{1}{r}{Male/Female} & 24 / 22 & 11 / 9 & 1 / 0 & 0 / 21 & 66 / 55 & 14 / 9 & 116 / 116 \\
\multicolumn{1}{r}{Right/Left eyes} & 46 / 0 & 20 / 0 & 1 / 1 & 21 / 21 & 117 / 115 & 22 / 21 & 227 / 158 \\
\multicolumn{1}{r}{Age (mean (SD))} & 47.5 (12.3) & 21.4 (2.3) & 23.0 (0.0) & 32.8 (5.4) & 50.8 (5.6) & 21.8 (7.9) & 42.9 (13.7) \\
\multicolumn{1}{l}{Device manufacturer} & Heidelberg & Heidelberg & Heidelberg & Heidelberg & Heidelberg & Topcon & All \\
\multicolumn{1}{l}{Device type} & Standard & Standard & FLEX & FLEX & Standard & DRI Triton Plus & All \\
\multicolumn{1}{l}{Scan location} &  &  &  &  & & & \\
\multicolumn{1}{r}{Horizontal/Vertical} & 168 / 0 & 55 / 50 & 4 / 4 & 76 / 76 & 381 / 369 & 132 / 139 & 816 / 638 \\
\multicolumn{1}{r}{Volume/Radial/Peripapillary} & 0 / 0 / 0 & 0 / 0 / 66 & 365 / 0 / 0 & 2,408 / 0 / 0 & 0 / 0 / 0 & 0 / 1,307 / 0 & 2,773 / 1,307 / 66 \\
\multicolumn{1}{r}{Total B-scans} & 168 & 171 & 373 & 2,560 & 750 & 1,578 & 5,600\\
\bottomrule
\end{tabular}

}
\end{adjustbox}
\caption{Overview of population characteristics. SD, standard deviation. Note that one participant's sex from the GCU Topcon cohort was not recorded.}
\label{tab:demo_tab}
\end{table*}

\subsection{Study population}
Our dataset contains 5,600 OCT B-scans of 233 participants from 6 cohorts of healthy and diseased individuals, unrelated to ocular pathology; \textbf{OCTANE}\cite{dhaun2014optical}, a longitudinal cohort study investigating choroidal microvascular changes in renal transplant recipients and healthy donors; \textbf{Diurnal Variation}\cite{dhaun2014optical}, a sub-cohort of OCTANE of young individuals investigating the possible effects of diurnal variation on the relationship between the choroid and markers of renal function; \textbf{Normative}, a detailed OCT examination of one of the authors (J.B.) with informed consent; \textbf{i-Test}\cite{dhaun2014optical}, a cohort of pregnant women evaluating whether the choroidal microvasculature reflects cardiovascular changes in both healthy and complicated pregnancies; \textbf{Prevent Dementia}, a longitudinal cohort tracking middle-aged individuals with varying risk of developing late onset Alzheimer's dementia \cite{ritchie2012prevent}; \textbf{GCU Topcon}\cite{moukaddem2022comparison}, an investigation into diurnal variation of the choroid in emmetropic and myopic individuals. All studies adhered to the Declaration of Helsinki and received relevant ethical approval and informed consent from all subjects was obtained in all cases from the host institution. \cref{tab:demo_tab} describes the population statistics and image acquisition statistics for each cohort.

Three OCT device types were used from two device manufacturers: the spectral domain OCT SPECTRALIS Standard Module OCT1 system and the spectral domain OCT SPECTRALIS portable FLEX Module OCT2 system (both Heidelberg engineering, Heidelberg, Germany), and the swept source OCT DRI Triton Plus (Topcon, Tokyo, Japan). For the Heidelberg devices, active eye tracking with built-in Automatic Real Time (ART) software was used with horizontal and vertical line scans capturing a 30$^\circ$ (9mm) fovea-centred region of interest, with an ART of 100, i.e. each final B-scan is the average of 100 B-scans. Posterior pole macular line scans covered a 30-by-25-degree rectangular region of interest using 31 consecutive scans, each with an ART of 50 (Posterior pole scans in the Normative cohort were acquired with an ART of 9). All Heidelberg data was collected at a pixel resolution of $768 \times 768$ pixels, with a signal quality $\geq$ 15. The Topcon device imaged the macular region using 12 fovea-centred radial scans, spaced 30$^\circ$ apart and covering a 30$^\circ$ (9mm) region of interest. Each B-scan had a resolution of $992 \times 1024$ pixels which was cropped horizontally by 32 pixels and resized to the resolution of the Heidelberg scans of $768 \times 768$. All Topcon data had an image quality score > 88 determined by the built-in TopQ software.

Five of the six cohorts were split into training (4,144 B-scans, 122 subjects), validation (466 B-scans, 28 subjects) and internal test sets (756 B-scans, 37 subjects) containing approximately 75, 10, and 15\% of the B-scans. We split the data on the subject-level, such that no individual ended up in more than one set. The remaining cohort, OCTANE, was entirely held-out as an external test set (168 B-scans, 46 individuals). Supplementary \cref{supp_tab:trainvaltest_population_overview_tab} gives a detailed overview of population and image characteristics for each of the four sets.
\subsection{Ground-truth (GT) labels}

The fovea coordinate was defined as the horizontal (column) pixel index which aligned with the deepest point of the foveal pit depression \cite{xuan2023deep}, i.e. where the central foveal pit was most illuminated, typically aligning with a ridge formed at the photoreceptor layer. The choroidal region was defined as the space posterior to the boundary delineating the retinal pigment epithelium layer and Bruch's membrane complex (RPE-Choroid, RPE-C, junction) and superior to the boundary delineating the sclera from the posterior most point of Haller's layer (Choroid-Sclera, C-S, junction). Between the choroid and sclera lies the suprachoroidal space, which is rarely visible on OCT B-scans and we consider not to be part of the choroid itself. The choroidal space is made up of interstitial fluid, or stroma, seen as brightly illuminated strips in the OCT B-scans, with interspersed, irregular areas of darker intensity representing choroidal vasculature This has been both empirically observed \cite{sohrab2012pilot, branchini2013analysis} and widely accepted among the research community \cite{agrawal2020exploring}. The choriocapillaris, a dense network of choroidal capillaries, is seen as a small band below Bruch's membrane complex approximately 10 microns thick \cite{nickla2010multifunctional} (roughly 3 pixels deep in OCT B-scans), and is assumed as part of the choroidal vasculature alongside larger vessels seen in Haller and Sattler's layers.

For OCT B-scans centred at the fovea (i.e. horizontal, vertical and radial scans), the foveal column location was detected manually. Those not centred at the fovea do not show the fovea. The GTs for choroidal region segmentation were generated using DeepGPET \cite{burke2023opensource} with the default threshold of 0.5. 897 scans were excluded from the dataset (and removed from \cref{tab:demo_tab} and supplementary \cref{supp_tab:trainvaltest_population_overview_tab}) because of poor region segmentations --- these were primarily Topcon B-scans which DeepGPET had not been trained on before.

GTs for vessel segmentation were generated using a novel, multi-scale quantisation and clustering-based approach, called multi-scale median cut quantisation (MMCQ), which we found to produce superior results to standard application of Niblack in preliminary analysis on the training set. MMCQ segments the choroidal vasculature by performing patch-wise local contrast enhancement at several scales using median cut clustering (quantisation) \cite{heckbert1982color} and histogram equalisation. The pixels of the subsequently enhanced choroidal space are then clustered globally using median cut clustering once more, classifying the pixels belonging to the clusters with the darkest pixel intensities as vasculature. The code for this algorithm is freely available here [LINK TO BE ADDED UPON ACCEPTANCE]. 

To improve the fidelity and robustness of our vessel segmentation GTs, we randomly varied the brightness and contrast of each OCT B-scan before application of MMCQ. We used 5 linearly spaced gamma levels to fix the mean brightness of each image between 0.2 and 0.5 and simultaneously altered the contrast using 5 linearly spaced factors between 0.5 and 3. A 3:2 majority vote for vessel label classification was used across all 25 variants. This improves robustness as spurious over- and undersegmentation contigent on specific image statistics are averaged out.

\subsection{Choroidalyzer's deep learning model}
Choroidalyzer segments the choroid region and vessels, and detects the fovea using a UNet deep learning model \cite{ronneberger2015u} with a depth of 7. This relatively high depth allows our model to better consider the global context. The first 3 blocks increase the internal channel dimension from 8 to 64, after which it is kept constant to reduce memory consumption and parameter count. Blocks consist of two convolutional layers, each followed by BatchNorm \cite{ioffe2015batch} and ReLU activation. Our up-blocks use a $1\times1$ convolution to reduce the channel dimension followed by bilinear interpolation, which is more compute and memory efficient than the standard transposed convolutions. We train our model for 40 epochs using the AdamW optimizer \cite{loshchilov2017decoupled} with a learning rate of $5\times10^{-4}$ and weight decay of $10^{-8}$ to minimise binary crossentropy, clamping the maximum gradient norm to $3$ before each step. We use automatic mixed precision to speed up training dramatically while reducing memory consumption by almost half. Forward pass and loss computation are done in bfloat16, a half-precision datatype optimised for machine learning.

During training, we apply the following data augmentations in random order per sample: Horizontal flip ($p=0.5$), changing the brightness and contrast independently (factors $\sim U(0.5, 1.5)$, $p=0.95$), random rotation and shear (degrees $\sim U(-25, 25)$ and $\sim U(-15, 15)$ respectively, $p=\nicefrac{1}{3}$), scaling the image (factor $ \sim U(0.8, 1.2)$, $p=\nicefrac{1}{3}$), where $U(a,b)$ denotes a uniform distribution between $a$ and $b$, and $p$ the probability of the transform being applied. For peripapillary scans which have a resolution of $1536\times768$, we use a crop of $768\times768$ using a random multiple of $192$ as offset per example and epoch.

The fovea is only a single point which would be difficult for a segmentation model to learn, as predicting close to 0 for all pixels would yield virtually the same loss as a perfect prediction. Thus, we create a target 51 pixels high and 19 pixels wide centred at the GT fovea location. The exact fovea location is set to 1, the whole column to 0.95, and adjacent columns to $0.95-(d*0.1)$ where d is the column distance from the fovea. Finally, we employ one-sided label smoothing and set all other pixels to 0.01 instead of 0 to stabilise training. We extract fovea column predictions by applying a 21-width triangular filter to the column-wise sums of our model's predictions and taking the column with the highest value.

\subsection{Statistical analysis}
We evaluate agreement in segmentations using the area under the receiver operating characteristic curve (AUC) and Dice coefficient, applying a fixed threshold of 0.5 to binarize our model's predictions. For the fovea column location, we use mean absolute error (MAE) and median absolute error (Median AE). For derived choroid metrics, we evaluate agreement with Pearson and Spearman correlations and further report MAEs. 

All choroidal metrics were computed using a region of interest (ROI) centred at the foveal pit, measuring 3mm temporally and nasally --- the ROI for volume scans was centred at the middle column index of the image --- corresponding to the standardised ROI according to the Early Treatment Diabetic Retinopathy Study (ETDRS) macular grid of 6,000 $\times $ 6,000 microns \cite{early1991early}. As peripapillary scans do not allow for a fovea-centred region of interest, we only look at segmentation metrics and use a threshold of 0.25 for vessel predictions. Area was computed by counting the pixels within the ETDRS grid, while thickness was measured at three linearly spaced locations, spanning the ETDRS grid, as point-source micron distances between the RPE-C and C-S junctions, locally perpendicular to the RPE-C junction.

Choroid vascular index is the ratio of vessel to non-vessel pixels in the choroid within the ETDRS grid. Our deep learning model outputs probabilities instead of discrete predictions, which capture uncertainty. As capturing uncertainty is desirable, we propose a ``soft'' vascular index which takes the ratio of predicted probabilities instead of discretized binary predictions. On the validation set, we found that this improves agreement.

To examine and characterise the behaviour of our model, we analysed cases of high error in detail. Concretely, for each of the three tasks (region and vessel segmentation, fovea detection), we selected the 15 examples from each test set where Choroidalyzer produced the highest errors. For redundant cases (i.e. adjacent, highly-similar slices from a volume scan) only one was retained. For fovea detection, cases of low error were also discarded. This left 28 cases for region, 29 for vessel, and 25 for fovea.

An adjudicating clinical ophthalmologist (I.M.) was provided with the original image, Choroidalyzer's prediction and the GT while being masked to the identity of the methods. Images and labels were provided individually and as composites. For each example, the adjudicator was asked which label they preferred. They also rated each label qualitatively on a 5-level ordinal scale (``Very bad'', ``Bad'', ``Okay'', ``Good'' and ``Very good'') for region segmentation quality, as well as intravascular and interstitial vessel segmentation quality. The latter two to quantify any potential under-segmentation of vessels and over-segmentation of the interstitial space.

Finally, we selected a random subsample of 20 B-scans at the patient-level from the external test set to be manually segmented by two graders, M1 and M2. M1 was a clinical ophthalmologist (I.M.) and M2 was a a PhD student who has worked with choroidal OCT data for the last 4 years (J.B.). Manual graders segmented the region and choroidal vessels using ITK-Snap \cite{py06nimg}. The manual segmentations were compared to Choroidalyzer and to the current state-of-the-art, namely DeepGPET for region segmentations \cite{burke2023opensource} and Niblack for vessel segmentation using a window size of 51 and standard deviation offset of -0.05, which mirrors previously published work \cite{muller2022application}.

\section{Results}

\begin{table*}[tb]
\centering
\begin{adjustbox}{max width=\textwidth}
\begin{tabular}{@{}llllllllllllllll@{}}
\toprule
  \multicolumn{1}{c}{\multirow{2}{*}{Set}} &

  \multicolumn{2}{c}{Region} &
  \multicolumn{2}{c}{Vessel} &
  \multicolumn{2}{c}{Fovea} &
  \multicolumn{3}{c}{Thickness} &
  \multicolumn{3}{c}{Area} &
  \multicolumn{3}{c}{Vascular Index}\\ 
  \cmidrule(l){2-3}\cmidrule(l){4-5}\cmidrule(l){6-7}\cmidrule(l){8-10}\cmidrule(l){11-13}\cmidrule(l){14-16}
  \multicolumn{1}{c}{} &
  \multicolumn{1}{c}{AUC} &
  \multicolumn{1}{c}{Dice} &
  \multicolumn{1}{c}{AUC} &
  \multicolumn{1}{c}{Dice} &
  \multicolumn{1}{c}{MAE} &
  \multicolumn{1}{c}{Median AE} &
  \multicolumn{1}{c}{pearson} &
  \multicolumn{1}{c}{spearman} &
  \multicolumn{1}{c}{MAE ($\mu$m)} &
  \multicolumn{1}{c}{pearson} &
  \multicolumn{1}{c}{spearman} &
  \multicolumn{1}{c}{MAE (mm$^2$)} &
  \multicolumn{1}{c}{pearson} &
  \multicolumn{1}{c}{spearman} &
  \multicolumn{1}{c}{MAE}\\ \midrule
Internal test & 0.9998 & 0.9789 & 0.9982 & 0.8817 & 3.9 px & 3.0 px & 0.9754$^{***}$ & 0.9692$^{***}$ & 8.2252 & 0.9815$^{***}$ & 0.9786$^{***}$ & 0.0385 & 0.8285$^{***}$ & 0.8097$^{***}$ & 0.0206 \\
External test & 0.9998 & 0.9749 & 0.9980 & 0.8703 & 3.4 px & 3.0 px & 0.9831$^{***}$ & 0.9868$^{***}$ & 8.0888 & 0.9779$^{***}$ & 0.9848$^{***}$ & 0.0487 & 0.7948$^{***}$ & 0.7991$^{***}$ & 0.0306\\ \bottomrule
\end{tabular}%

\end{adjustbox}
\caption{Metrics for Choroidalyzer against ground-truth annotations from the internal and external test sets. $^{***}$ indicates p<0.0001.}
\label{tab:main_results}
\end{table*}

\begin{figure*}[tb]
\centering
\includegraphics[width=\textwidth]{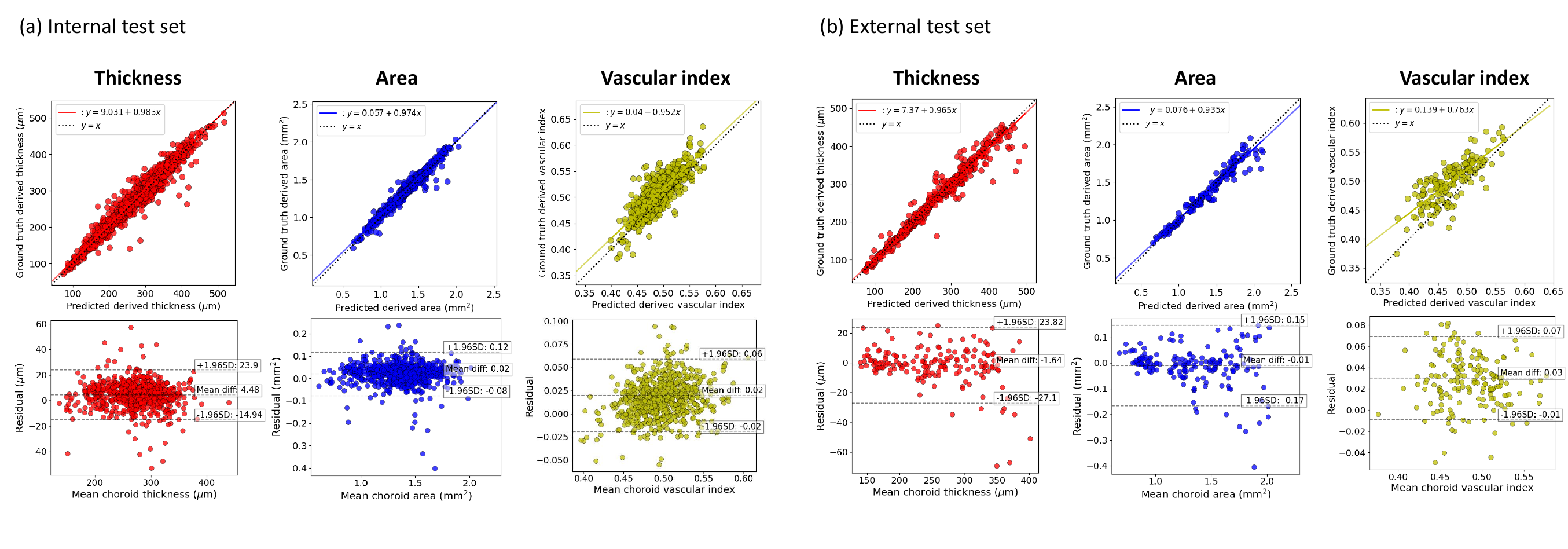}
\caption{Agreement in thickness, area and vascular index for a) the internal and b) the external test sets. Top row are scatterplots with best regression fit and identity lines, bottom row are Bland-Altman plots. Note that we chose to fit each plot to the data range and thus the scale of the axes are not exactly the same between internal and external test sets, especially for vascular index.}
\label{fig:corr_plot}
\end{figure*}

\subsection{Performance on internal and external test sets}

\cref{tab:main_results} shows the performance of Choroidalyzer on the internal and external test sets. Our model achieves very good performance in terms of AUC and Dice for region and vessels on both sets. Metrics for region are higher than for vessels, which is expected as choroidal vessel segmentation is much more difficult and ambiguous than region segmentation, and thus the GTs are themselves imperfect. Performance was slightly higher for the internal test set than the external test set, which is expected, but only marginally so, indicating that our model generalises well to new cohorts. For the peripapillary scans which only exist in the internal test set, our model achieved an AUC of 0.9996 (region) / 0.9925 (vessel) and Dice of 
0.9636 (region) / 0.7155 (vessel). This is reasonable performance but lower than for other scans. 

\begin{figure*}[tb]
\centering
\includegraphics[width=\textwidth]{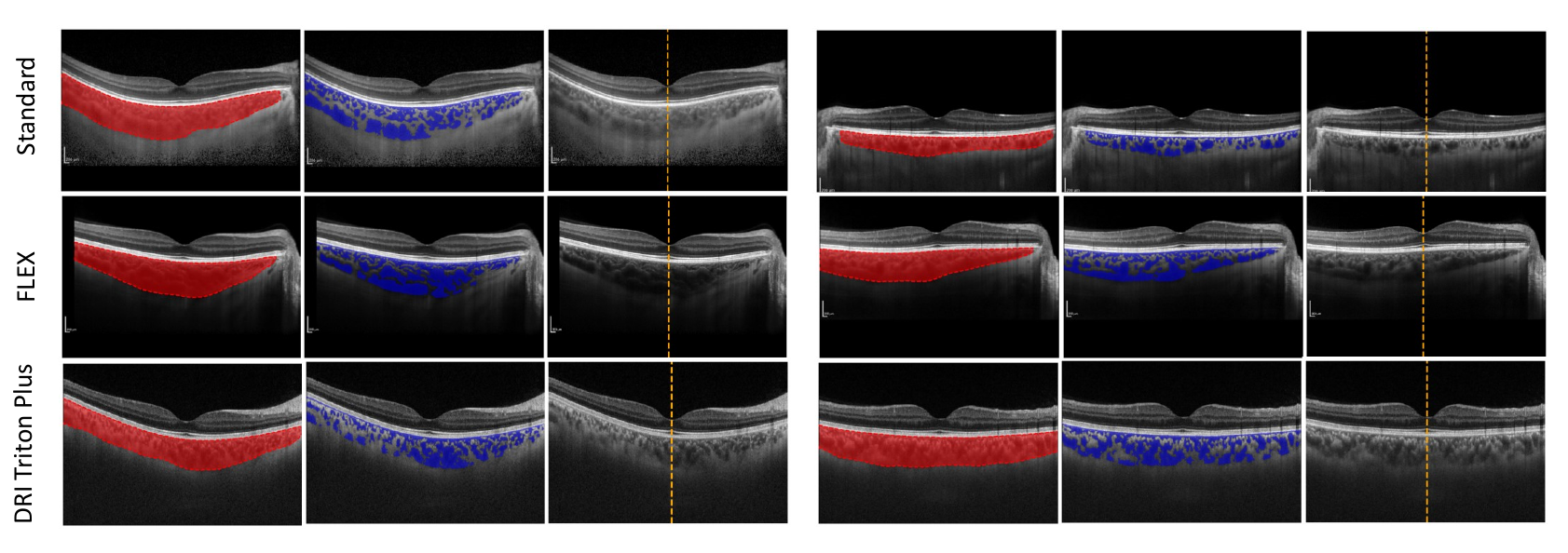}
\caption{Examples of Choroidalyzer being applied to scans from different imaging devices. Six fovea-centred OCT B-scans, two per imaging device type, from the internal test set showing region segmentations (left), vessel segmentations (middle), and fovea column location (right).}
\label{fig:choroidalyzer_examples}
\end{figure*}

For fovea detection, the model had a MAE of 3.9 px for the internal and 3.4 px for the external test set, with the median absolute error being 3 px for both. This is excellent performance as an error of 3 px on a 768 px-wide image will not meaningfully change our region-of-interest or resulting metrics (data not shown --- see the supplementary materials for the analysis effects of fovea location on downstream metrics). For the derived choroid metrics, Choroidalyzer shows excellent agreement with the GTs on thickness and area, with Pearson and Spearman correlations of 0.9692 or greater for both internal and external test sets. For the vascular index, performance is a bit lower, with correlations between 0.7948 and 0.8285. Although vascular index depends on both region and vessel segmentation, the other metrics indicate that the differences in vascular index are driven primarily by differences in vessel segmentation. Still, the observed correlations are high in absolute terms. \cref{fig:corr_plot} shows correlation and Bland-Altman plots for the three derived metrics on both test sets, which likewise indicate generally very good agreement. \cref{fig:choroidalyzer_examples} shows some examples for each of the three imaging devices.

\subsection{Comparison with manual segmentations}
\cref{tab:manual_comparison} shows the results from manual segmentations. For automated methods, we compare with each manual grader and then averaged the performance across both graders to make the results more concise. The comparisons with individual graders are reported in the supplementary  \cref{supp_tab:man_compar_full}. Interestingly, while vessel Dice for Choroidalyzer (0.7410 vs M1 and 0.7927 vs M2; mean 0.7669) is again much worse than region Dice and even worse than the vessel Dice on both test sets, it is very similar to the inter-grader agreement of 0.7699. More generally, the inter-grader agreements for all other metrics are similar to Choroidalyzer's agreement with the graders, with the notable exception of vascular index. Here, Choroidalyzer's MAE is better (0.0555 vs M1 and 0.0506 vs M2; mean 0.0531) than the inter-grader agreement (0.0618), as is the Spearman correlation, but Pearson correlation and ICC are worse. Compared to the respective state-of-the-art (SOTA, i.e. DeepGPET for region, Niblack for vessel segmentation), Choroidalyzer has better agreement with the graders for most of the metrics, although methods are generally comparable.

\cref{tab:manual_comparison_times} shows the time per scan for the manual graders and automatic approaches. The manual graders on average needed more than 26 and 22 minutes (mean 24), with the vast majority of that time spent on the vessel segmentation. By contrast, the automatic methods on a standard laptop needed about a second per scan and no human time at all. Thus, to get through a dataset of 100 scans, it would take manual graders about 40 hours of work, but with automated methods it would be less than 2 minutes. With GPU-acceleration, Choroidalyzer and DeepGPET could achieve throughputs of dozens or hundreds of scans per second even on consumer-grade hardware. Comparing the automated methods with each other, Choroidalyzer took 73\% less time than DeepGPET and Niblack, while also detecting the fovea location. All three methods are very fast but for very large datasets or deployment on edge devices, Choroidalyzer's efficiency is an additional advantage over existing automated methods.

\begin{table*}[tb]
\centering
\begin{adjustbox}{max width=\textwidth}
{\small

\begin{tabular}{@{}lllllllllllllllll@{}}
\toprule
  \multicolumn{1}{c}{\multirow{2}{*}{Comparison}} &
  \multicolumn{2}{c}{Region} &
  \multicolumn{2}{c}{Vessel} &
  \multicolumn{4}{c}{Thickness} &
  \multicolumn{4}{c}{Area} &
  \multicolumn{4}{c}{Vascular Index}\\ 
  \cmidrule(l){2-3}\cmidrule(l){4-5}\cmidrule(l){6-9}\cmidrule(l){10-13}\cmidrule(l){14-17} 
  \multicolumn{1}{c}{} &
  \multicolumn{1}{c}{AUC} &
  \multicolumn{1}{c}{Dice} &
  \multicolumn{1}{c}{AUC} &
  \multicolumn{1}{c}{Dice} &
  \multicolumn{1}{c}{pearson} &
  \multicolumn{1}{c}{spearman} &
  \multicolumn{1}{c}{ICC} &
  \multicolumn{1}{c}{MAE ($\mu$m)} &
  \multicolumn{1}{c}{pearson} &
  \multicolumn{1}{c}{spearman} &
  \multicolumn{1}{c}{ICC} &
  \multicolumn{1}{c}{MAE (mm$^2$)} &
  \multicolumn{1}{c}{pearson} &
  \multicolumn{1}{c}{spearman} &
  \multicolumn{1}{c}{ICC} &
  \multicolumn{1}{c}{MAE}\\ \midrule
M1 vs. M2 & 0.9639 & 0.9474 & 0.8891 & 0.7699 & 0.9503 & 0.9521 & 0.9783 & 17.8833 & 0.9516 & 0.9248 & 0.9751 & 0.1096 & 0.8074 & 0.6857 & 0.8172 & 0.0618 \\
Choroidalyzer vs. Manual (avg) & 0.9978 & 0.9375 & 0.9914 & 0.7669 & 0.9534 & 0.9663 & 0.9873 & 20.9750 & 0.9554 & 0.9368 & 0.9756 & 0.1150 & 0.6654 & 0.7383 &  0.7613 & 0.0530 \\
SOTA vs. Manual (avg) & 0.9444 & 0.9333 & 0.9223 & 0.7742 & 0.9676 & 0.9636 & 0.9802 & 19.9250 & 0.9548 & 0.9233 & 0.9769 & 0.1202 & 0.6907 & 0.6105 & 0.7103 & 0.1682 \\ \bottomrule
\end{tabular}%

}
\end{adjustbox}
\caption{Comparison metrics for the 20 images assessed manually and algorithmically from the external test set. Comparisons made between the 2 manual graders, the proposed model and current state of the art, DeepGPET for region segmentation and the Niblack thresholding algorithm for vessel segmentation. SOTA: current state-of-the-art, i.e. DeepGPET for region and Niblack for vessel segmentation.}
\label{tab:manual_comparison}
\end{table*}

\begin{table}[tb]
\centering
\begin{adjustbox}{max width=0.5\textwidth}
{\small \begin{tabular}{@{}lccc@{}}
\toprule
\hspace{1em}Method\hspace{1.3em} & \hspace{1em}Region (s) \hspace{1.3em} & \hspace{1em} Vessel (s) \hspace{1.3em} & \hspace{1em} Total (s) \hspace{1.3em}\\ \midrule
M1 & 78.400 $\pm$ 12.261 & 1506.000 $\pm$ 744.073 & 1584.400 $\pm$ 771.284  \\
M2 & 165.000 $\pm$ 23.889 & 1176.700 $\pm$ 744.073 & 1341.700 $\pm$ 265.800   \\
SOTA & 0.751 $\pm$ 0.081 & 0.370 $\pm$ 0.105 & 1.121 $\pm$ 0.140\\
Choroidalyzer & - & - & 0.299 $\pm$ 0.018 \\
\bottomrule
\end{tabular}%


}
\end{adjustbox}
\caption{Mean (standard deviation) execution time of the four different approaches to region and vessel segmentation for the 20 images assessed manually and algorithmically from the external test set. SOTA: current state-of-the-art, i.e. DeepGPET for region and Niblack for vessel segmentation. Automated methods were run on a standard laptop with a 4 year old i5 CPU and 16 GB of RAM but no GPU.}
\label{tab:manual_comparison_times}
\end{table}

\subsection{Detailed error analysis}

\begin{table*}[tb]
\centering
\begin{adjustbox}{max width=\textwidth}
{\small 

\begin{tabular}{@{}lccc@{}}
\toprule
 & Preferred Choroidalyzer & Preferred SOTA & Both equally good \\
\midrule
Region & 8/28 & 5/28 & 15/28 \\
Vessel & 13/29 & 4/29 & 12/29 \\
Fovea & 23/25 & 1/25 & 1/25\\ 
\midrule
& & & \\
\midrule
Method & Region: quality & Vessel: intravascular & Vessel: interstitial \\ \midrule
Choroidalyzer & VG: 3, G: 14, O: 9, B: 1, VB: 1 & VG: 0, G: 17, O: 12, B: 0, VB: 0 & VG: 0, G: 20, O: 9, B: 0, VB: 0 \\
SOTA & VG: 0, G: 17, O: 8, B: 1, VB: 2 & VG: 0, G: 5, O: 19, B: 3, VB: 2 & VG: 0, G: 17, O: 8, B: 2, VB: 2 \\ 
\bottomrule
\hspace{0.5em}
\end{tabular}%

}
\end{adjustbox}
\caption{Preference and segmentation scores from masked expert adjudicator (I.M.) comparing the highest region segmentation, vessel segmentation and fovea column errors between Choroidalyzer and the ground truth labels. VG, very good; G, good; O, okay; B, bad; VB, very bad.}
\label{tab:ian_adj}
\end{table*}

\cref{tab:ian_adj} shows the results of manual inspection of scans where Choroidalyzer produced the highest error compared to the GT on the test sets. For region segmentation, Choroidalyzer was preferred in 8 cases, the GT in 5, and both methods were considered equally good in 15 cases. In terms of quality, Choroidalyzer was ``Very bad'' in only one case compared with 2 for the GT and ``Very good'' 3 times compared to none for the GT. For the vessels, Choroidalyzer was preferred in 13 cases, the GT in 4, and both were tied in 12 cases. Vessel segmentation is a harder task, with no methods achieving ``Very good''. However, the  intravascular scores for Choroidalyzer are substantially better, with no ``Bad'' or ``Very bad'' (vs. 3 and 2, respectively for GT) and far more ``Good'' (17 vs. 5), and the interstitial scores are similarly better. Finally, for the fovea, Choroidalyzer was preferred 23/25 times and the GT only twice, indicating that large fovea errors are almost exclusively due to mistakes in the manual GT labels.

\begin{figure*}[tb]
\centering
\includegraphics[width=\textwidth]{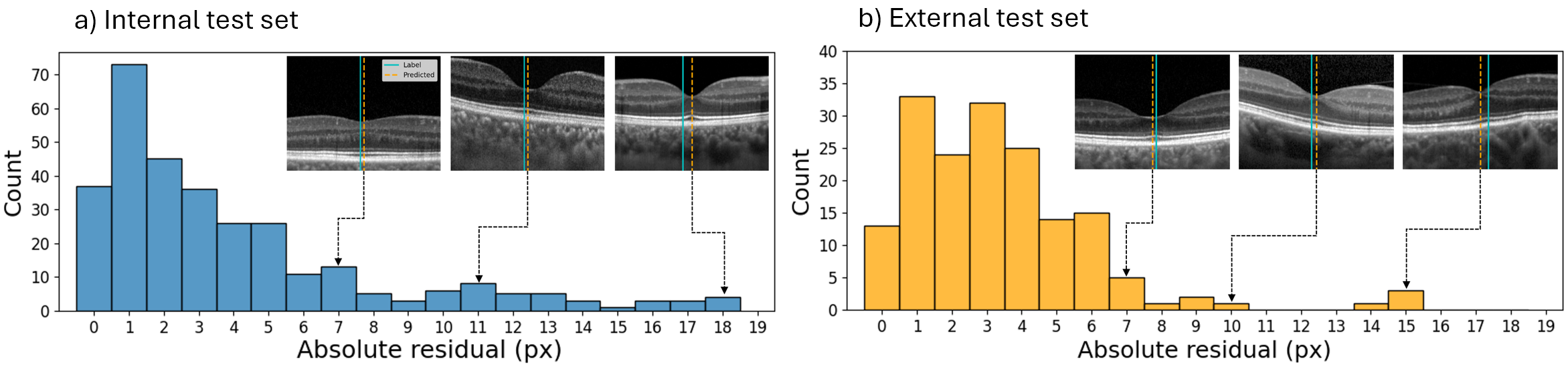}
\caption{Histogram of absolute errors for fovea column detection for the internal (left) and external (right) test sets. Examples for different levels of error are shown with dotted lines indicating which part of the distribution they come from. In the examples, the teal line indicates the GT label, the dashed orange line the prediction.}
\label{fig:fov_plot}
\end{figure*}

\cref{fig:fov_plot} shows the distributions of fovea errors for both test sets along with each example in both sets. For very large residuals (10+px), the GTs are wrong and Choroidalyzer correctly identifies the fovea location. For errors around 7px, still twice the MAE, both methods are similar with either method sometimes being more correct. Further exploration revealed the majority of incorrectly labelled ground-truths to be Topcon OCT B-scans, as each 12-stack of radial scans are not centred at the fovea, and initial manual annotation detected the fovea for only one to represent each stack. Despite this oversight, Choroidalyzer learned to dectect the fovea robust and accurately.

\begin{figure}[tb]
\centering
\includegraphics[width=0.5\textwidth]{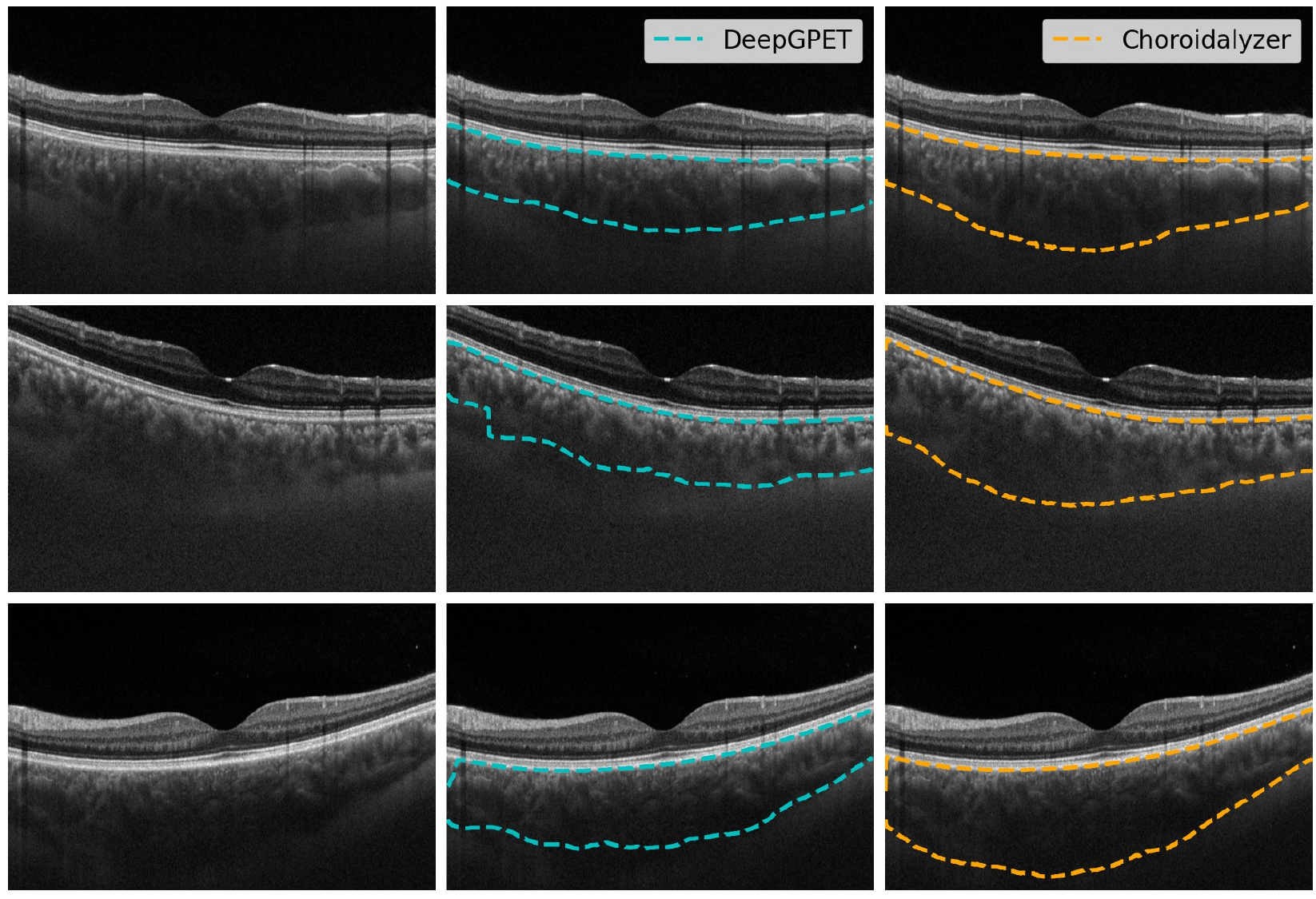}
\caption{Three example Topcon OCT B-scans with successful region segmentations from Choroidalyzer (right) and failed segmentations from DeepGPET (middle).}
\label{fig:Topcon_bad_}
\end{figure}

\section{Discussion}
We developed Choroidalyzer, an end-to-end pipeline for choroidal analysis. Choroidalyzer shows excellent performance on the internal and external test sets. Where Choroidalyzer produced the highest errors were primarily cases of imperfect GTs and Choroidalyzer was generally preferred by a blinded adjudicating ophthalmologist (I.M.), further indicating robustness and good performance. Its agreement with manual segmentations, which demand substantial time and attention from a human expert, is comparable to the inter-grader agreement. This suggests that Choroidalyzer performs well compared to laborious manual segmentation and also highlights the subjectivity introduced by manual graders. Choroidalyzer not only produces results similar to that of a skilled manual grader, it also does so fully-automatically without introducing subjectivity and thus increases standardisation and reproducibility. If researchers use Choroidalyzer, their results are repeatable and would be much more comparable to other studies also using Choroidalyzer than if different manual graders were used in each case. 

Additionally, Choroidalyzer saves a substantial amount of time per image over manual segmentation, freeing up researcher time and enabling large scale analyses that otherwise would not be possible. Even compared to the current state-of-the-art for automated methods, DeepGPET and Niblack, Choroidalyzer can do the analysis in roughly a quarter of the time. More importantly, Choroidalyzer provides an end-to-end pipeline which makes it easier to implement and use than having to combine multiple methods like Niblack and DeepGPET. Ease-of-use is often underappreciated in the literature but key in saving researchers time and allowing them to focus on the science.

Choroidalyzer performed well against manual graders relative to the state of the art methods, reaching or surpassing the levels of agreement even between the two manual graders, particularly for vascular index, a far more difficult metric to calculate accurately than area and thickness. The inter-grader agreement between manual graders for these metrics indicate a potential lower bound of what effect sizes we might expect from these metrics. This has important downstream impact on the statistical confidence of results from cohort studies, particularly when assessing the choroidal vasculature \cite{wei2018comparison, agrawal2016choroidal}. 

It is often difficult to visualise the choroid due to imaging noise, poor eye tracking and patient fixation, or operator inexperience. Thus, in some cases vessel boundaries can be hard to discern. This is why we proposed to use a soft version of choroid vascular index where the probabilities that Choroidalyzer outputs are used instead of thresholded, binarized segmentations. The probabilities capture uncertainty about the precise location of the vessel wall and thus is more robust than using a single, somewhat arbitrary threshold. Users could also tune the binarization threshold for their own images, if desired, which might help in instances of poor visibility of the choroidal vasculature.

Segmentation performance for peripapillary scans was reasonable but much worse than for other scan types. This could be due to those scans being relatively rare in our dataset and showing parts of the retina on the nasal side of the optic disc that are not captured in fovea-centred scans. More peripapillary training data would likely increase performance. In our opinion, at present Choroidalyzer can be used for these scans but requires subsequent manual inspection and potential correction. Furthermore, adjusting the binarization threshold for the vessel predictions can improve results.

Our model detected the fovea well, and the largest errors were cases where ground-truths were incorrectly labelled with the model correctly identifying the fovea location as confirmed by masked adjudication. Thus, the model performed even better than what the quantitative results suggest. In the present work, we have focused on identifying the fovea column which is needed to define the fovea-centred region of interest. However, after selecting and evaluating our final model, we realised that in relatively rare, highly myopic cases, the retina and choroid can be at a steep angle relative to the image axes. For those, it would be best to define the region of interest along the choroid axes rather than image axes, most easily done by drawing a centre line from the fovea perpendicularly through the retina and choroid. Thus, it could be useful to also segment the retina and to determine both the row and column of the fovea. While not our initial objective, we did some preliminary analyses and found that we can derive the fovea row well with our current model (data not shown). Furthermore, to understand the effect of fovea location error on downstream choroidal metrics, we simulated random per-sample deviations of $\pm6$px, twice the Median AE, and found that they yielded virtually identical results (see supplementary \cref{supp_tab:fovea_pearson_boxplots} and \cref{supp:fovea_corr_ba_plots}).

The dataset in the present work was substantially larger than the one used for DeepGPET and importantly contains both Heidelberg and Topcon scans. As a result, Choroidalyzer can segment even difficult Topcon scans where DeepGPET failed (\cref{fig:Topcon_bad_}). Choroidalyzer was trained on region and vessel GTs generated by fully- and semi-automatic methods, respectively, which were then checked for errors and only manually improved where needed. Recent work argues that such approaches to generating GTs are preferable as they reduce subjectivity and thus bias and inconsistency\cite{maloca2023human}.

Choroidalyzer also has limitations. Most importantly, there is no quality scoring component to reject B-scans that do not show the choroid in sufficient detail to allow for reasonable analysis. While modern OCT devices typically show the choroid in good detail, especially if EDI is used, this is not always the case. Most devices provide some quality indicators, but we have not investigated quality thresholds for specific devices, below which Choroidalyzer would not function. Furthermore, OCT quality indicators are typically focused on the retina, and although poor visualisation of the retina might imply poor visualisation of the choroid, the reverse is not necessarily the case. A quality scoring method specific to the choroid would be a useful addition to the field. Another limitation is that Choroidalyzer was trained only on cohorts relating to systemic health but not ocular disease or data acquired during routine clinical practice.

Future work could improve the underlying deep learning model of Choroidalyzer, e.g. by training and evaluating it on data from more diverse sources. Data with ocular pathology, e.g. abnormally sized choroids due to myopia, age-related macular degeneration or central serous chorio-retinopathy, could be used to investigate whether Choroidalyzer is robust in those contexts and to train an improved version if needed. Moreover, automated quality scoring methods relating to the choroid would address a key need in choroidal analysis. Finally, Choroidalyzer could be extended to measure additional choroidal metrics, such as macular thickness and vessel density maps across a volume, or relating to its curvature.

\section{Conclusion}
Choroidal thickness, area, and especially vascular index are highly interesting metrics and potential biomarkers for both systemic and ocular health. However, calculating them used to be laborious and - when done manually - subjective. Choroidalyzer provides an efficient, end-to-end pipeline to alleviate these problems. We hope that by making Choroidalyzer openly accessible, we will enable researchers and clinicians to conveniently calculate these metrics and use them for their research, while improving reproducibility and standardisation in the field.

\FloatBarrier

\section*{Acknowledgements}
J.E. was supported by UK Research and Innovation (grant EP/S02431X/1) as part of the Centre of Doctoral Training in Biomedical AI at the School of Informatics, University of Edinburgh. J.B. was supported by the Medical Research Council (grant MR/N013166/1) as part of the Doctoral Training Programme in Precision Medicine at the Usher Institute, University of Edinburgh.

For the purpose of open access, the authors have applied a creative commons attribution (CC BY) licence to any author accepted manuscript version arising. 

The authors acknowledge the support provided by the Edinburgh Imaging and Edinburgh Clinical Research Facility at the University of Edinburgh, and thank all participants in the studies used in this paper. Supported in part by the Alzheimer's Drug Discovery Foundation (project no. GDAPB-201808-2016196); NHS Lothian R\&D; British Heart Foundation Centre for Research Excellence Award III (RE/18/5/34216). Supported in part also by the Wellcome Leap In Utero scheme. The funding sources were not involved in designing, conducting, or submitting this work.

\section{Conflicts of Interest}
The authors declare no conflicts of interest.


\onecolumn
 
\setcounter{figure}{0}
\renewcommand{\thefigure}{S\arabic{figure}}
\setcounter{table}{0}
\renewcommand{\thetable}{S\arabic{table}}
\pagebreak

\section*{Supplementary Material}

\begin{table}[H]
\centering
{\small
\begin{tabular}{@{}lcccc|c@{}}
\toprule
 & Training & Validation & Testing & External test & Total \\
 \midrule
\multicolumn{1}{l}{Subjects} & 122 & 28 & 37 & 46 & 233 \\
\multicolumn{1}{r}{Male/Female} & 64 / 57 & 12 / 16 & 16 / 21 & 24 / 22 & 116 / 116 \\
\multicolumn{1}{r}{Control/Case} & 76 / 46 & 16 / 12 & 20 / 17 & 0 / 46 & 112 / 121 \\
\multicolumn{1}{r}{Right/Left eyes} & 117 / 107 & 27 / 23 & 37 / 28 & 46 / 0 & 227 / 158 \\
\multicolumn{1}{r}{Standard/FLEX/DRI Triton Plus} & 88 / 14 / 20 & 24 / 2 / 2 & 29 / 6 / 2 & 46 / 0 / 0 & 187 / 22 / 24 \\
\multicolumn{1}{r}{Heidelberg/Topcon} & 102 / 20 & 26 / 2 & 35 / 2 & 46 / 0 & 209 / 24 \\
\multicolumn{1}{r}{Age (mean (SD))} & 40.7 (14.2) & 42.5 (11.9) & 44.5 (13.4) & 47.5 (12.3) & 42.9 (13.4) \\
\multicolumn{1}{l}{Cohort} &  &  &  &  &  \\
\multicolumn{1}{r}{OCTANE} & 0 & 0 & 0 & 46 & 46 \\
\multicolumn{1}{r}{Diurnal Variation} & 12 & 4 & 4 & 0 & 20 \\
\multicolumn{1}{r}{Normative} & 1 & 0 & 0 & 0 & 1 \\
\multicolumn{1}{r}{i-Test} & 13 & 2 & 6 & 0 & 21 \\
\multicolumn{1}{r}{Prevent Dementia} & 76 & 20 & 25 & 0 & 121 \\
\multicolumn{1}{r}{GCU Topcon} & 20 & 2 & 2 & 0 & 24 \\
\multicolumn{1}{l}{B-scans} &  &  &  &  &  \\
\multicolumn{1}{r}{Standard/Flex/DRI Triton Plus} & 582 / 2,281 / 1,281 & 136 / 190 / 140 & 137 / 462 / 157 & 168 / 0 / 0 & 1,023 / 2,933 / 1,578 \\
\multicolumn{1}{r}{Heidelberg/TopCon} & 2,863 / 1,281 & 326 / 140 & 599 / 157 & 168 / 0 & 3,956 / 1,578 \\
&  &  &  &  & \\
\multicolumn{1}{r}{Horizontal/Vertical scans} & 462 / 461 & 90 / 82 & 95 / 95 & 168 / 0 & 816 / 638 \\
\multicolumn{1}{r}{Volume/Radial/Peripapilary scans} & 2,161 / 1,060 / 39  & 178 / 116 / 15 & 434 / 131 / 12 & 0 / 0 / 0 & 2,773 1,307 / 0\\
\multicolumn{1}{r}{Total B-scans} & 4,183 & 481 & 768 & 168 & 5,600 \\
\bottomrule
\end{tabular}
}
\caption{Overview of population and image characteristics of the internal training, validation and test sets, and also the external test set. Note that one participant's sex from the Topcon cohort was not recorded.}
\label{supp_tab:trainvaltest_population_overview_tab}
\end{table}

\begin{table*}[bh]
\centering
\begin{adjustbox}{max width=\textwidth}
{\small \begin{tabular}{@{}lllllllllllllllll@{}}
\toprule
  \multicolumn{1}{c}{\multirow{2}{*}{Comparison}} &
  \multicolumn{2}{c}{Region} &
  \multicolumn{2}{c}{Vessel} &
  \multicolumn{4}{c}{Thickness} &
  \multicolumn{4}{c}{Area} &
  \multicolumn{4}{c}{Vascular Index}\\ 
  \cmidrule(l){2-3}\cmidrule(l){4-5}\cmidrule(l){6-9}\cmidrule(l){10-13}\cmidrule(l){14-17} 
  \multicolumn{1}{c}{} &
  \multicolumn{1}{c}{AUC} &
  \multicolumn{1}{c}{Dice} &
  \multicolumn{1}{c}{AUC} &
  \multicolumn{1}{c}{Dice} &
  \multicolumn{1}{c}{pearson} &
  \multicolumn{1}{c}{spearman} &
  \multicolumn{1}{c}{ICC} &
  \multicolumn{1}{c}{MAE ($\mu$m)} &
  \multicolumn{1}{c}{pearson} &
  \multicolumn{1}{c}{spearman} &
  \multicolumn{1}{c}{ICC} &
  \multicolumn{1}{c}{MAE (mm$^2$)} &
  \multicolumn{1}{c}{pearson} &
  \multicolumn{1}{c}{spearman} &
  \multicolumn{1}{c}{ICC} &
  \multicolumn{1}{c}{MAE}\\ \midrule
M1 vs. M2 & 0.9639 & 0.9474 & 0.8891 & 0.7699 & 0.9503 & 0.9521 & 0.9783 & 17.8833 & 0.9516 & 0.9248 & 0.9751 & 0.1096 & 0.8074 & 0.6857 & 0.8172 & 0.0618 \\ \midrule
M1 & & & & & & & & & & & & & & & & \\
\multicolumn{1}{r}{Choroidalyzer} & \textbf{0.9964} & \textbf{0.9242} & \textbf{0.9896} & 0.7410 & 0.9322 & \textbf{0.9490} & \textbf{0.9761} & 27.2167 & \textbf{0.9211} & \textbf{0.8872} & 0.9570 & \textbf{0.1598} & \textbf{0.7668} & \textbf{0.8406} & \textbf{0.7265} & \textbf{0.0555} \\
\multicolumn{1}{r}{SOTA} & 0.9370 & 0.9227 & 0.9271 & \textbf{0.7714} & \textbf{0.9437} & 0.9378 & 0.9676 & \textbf{25.8500} & 0.9198 & 0.8692 & \textbf{0.9589} & 0.1631 & 0.7150 & 0.6857 & 0.7157 & 0.1901 \\ \midrule
M2 & & & & & & & & & & & & & & & & \\
\multicolumn{1}{r}{Choroidalyzer} & \textbf{0.9993} & \textbf{0.9507} & \textbf{0.9933} & \textbf{0.7927} & 0.9746 & 0.9838 & \textbf{0.9984} & 14.7333 & 0.9896 & \textbf{0.9865} & 0.9942 & \textbf{0.0702} & 0.5640 & \textbf{0.6361} & \textbf{0.7960} & \textbf{0.0506} \\ 
\multicolumn{1}{r}{SOTA} & 0.9175 & 0.9439 & 0.9175 & 0.7770 & \textbf{0.9914} & \textbf{0.9894} & 0.9927 & \textbf{14.0000} & \textbf{0.9897} & 0.9774 & \textbf{0.9948} & 0.0770 & \textbf{0.6663} & 0.5353 & 0.7047 & 0.1464 \\ \bottomrule
\end{tabular}%
}
\end{adjustbox}
\caption{Full comparison metrics between Choroidalyzer, two manual graders M1 and M2, and state of the art region and vessel segmentation methods DeepGPET and Niblack.}
\label{supp_tab:man_compar_full}
\end{table*}

\subsection{Analysis effects of fovea location error on downstream metrics}

\begin{figure*}[b]
    \centering
    \includegraphics[width=0.5\textwidth]{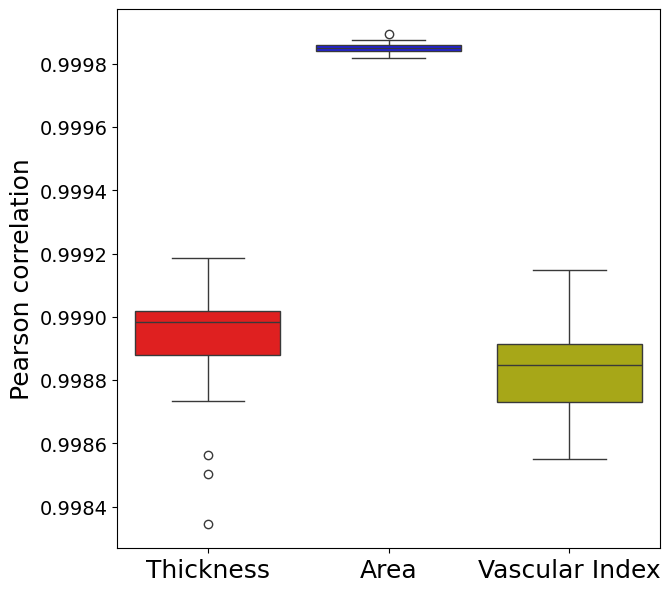}
    \caption{Distribution of Pearson correlation coefficients for each choroidal metric when perturbing the fovea coordinate column. Note the scale of the y-axis, even the lowest correlation we observed was $>0.99$.}
    \label{supp_tab:fovea_pearson_boxplots}
\end{figure*}

\begin{figure*}
    \centering
    \includegraphics[width=\textwidth]{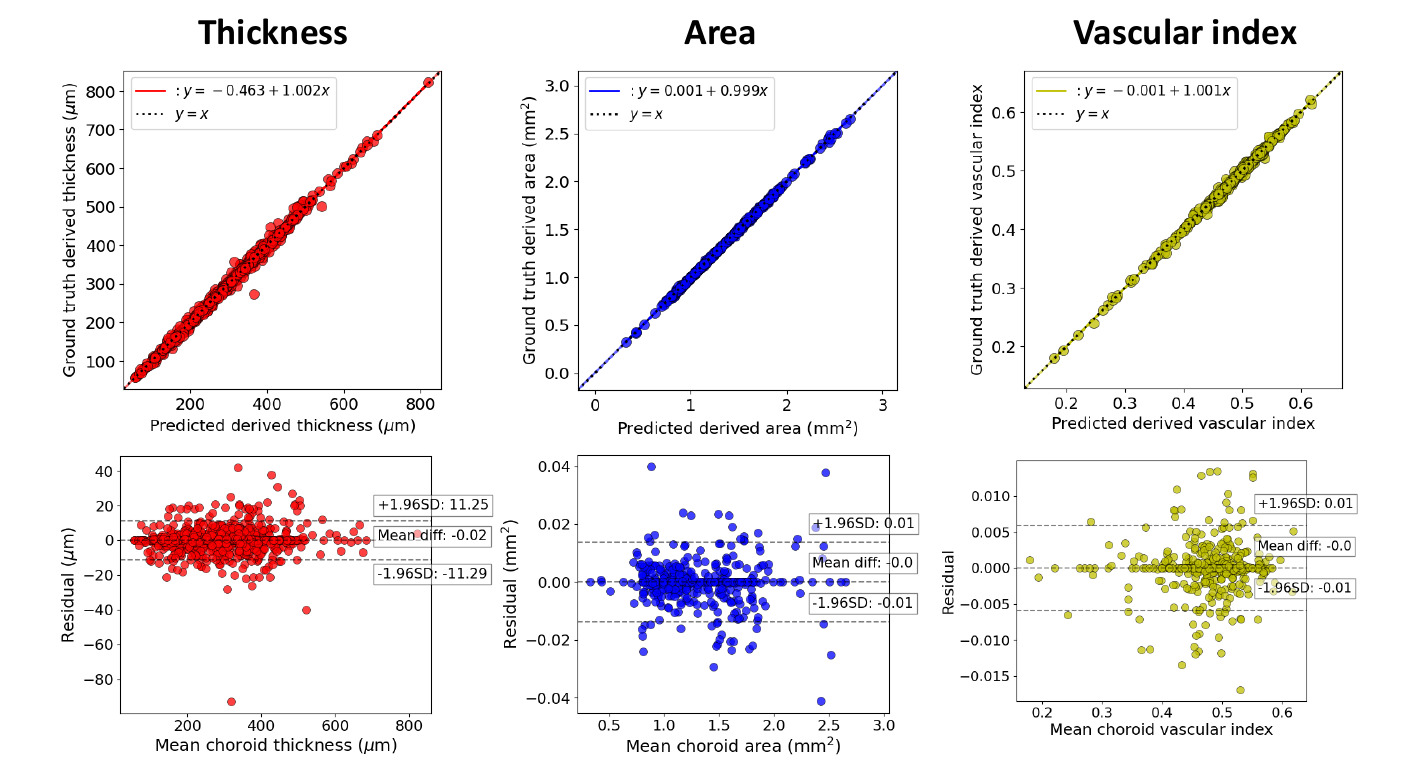}
    \caption{Correlation and Bland-Altman plots for choroidal metrics for the poorest performing simulation of perturbing the fovea column coordinate on a random 10\% subsample of the dataset.}
    \label{supp:fovea_corr_ba_plots}
\end{figure*}

Choroidalyzer measured the fovea column coordinate with a median absolute error of 3 pixels in both the internal and external test sets. We tested the effect of perturbing the fovea column on choroidal metrics by comparing fovea-centred metrics and metrics derived after the fovea column was randomly perturbed using a discretised uniform distribution $ \sim U(-6, 6\}$ (excluding 0). 50 simulations were run on approximately 10\% of the dataset (495 OCT B-scans), selected at random to represent eye type and location on the macula. All metrics had excellent Pearson correlation ($r$ > 0.99, p<0.00001, supplementary \cref{supp_tab:fovea_pearson_boxplots}). Scatterplots of metrics for the poorest performing simulation according to absolute error across all metrics (supplementary \cref{supp:fovea_corr_ba_plots}) shows excellent agreement with the identity line, with limits of agreement in the Bland-Altman plots well within acceptable bounds for all metrics \cite{rahman2011repeatability, agrawal2017influence}. Thus, the fovea column quantitative error observed from Choroidalyzer does not significantly impact the choroidal metrics.

\end{document}